\documentclass[useAMS,usenatbib]{mnras}
\usepackage{graphicx,natbib,times,amssymb}

\title[AGN from HeII]{Active Galactic Nuclei from \HeII: a more complete census of AGN in SDSS galaxies yields a new population of low-luminosity AGN in highly star-forming galaxies}
\author[Rudolf E. B{\"a}r et al.]{Rudolf E. B\"{a}r$^{1}$\thanks{E-mail: baerr@phys.ethz.ch}, Anna K. Weigel$^{1}$, Lia F. Sartori$^{1}$, Kyuseok Oh$^{1}$, Michael Koss$^{1}$\thanks{Ambizione fellow} \newauthor{and Kevin Schawinski$^{1}$}\\
$^{1}$Institute for Astronomy, Department of Physics, ETH Zurich, Wolfgang-Pauli-Strasse 27, CH-8093 Zurich, Switzerland}
\begin{document}

\def\Chandra{\textit{Chandra}}
\def\XMM{\textit{XMM-Newton}}
\def\Swift{\textit{Swift}}

% Forbidden Lines
\def\OI{[\mbox{O\,{\sc i}}]~$\lambda 6300$}
\def\OIII{[\mbox{O\,{\sc iii}}]~$\lambda 5007$}
\def\SII{[\mbox{S\,{\sc ii}}]~$\lambda \lambda 6717,6731$}
\def\NII{[\mbox{N\,{\sc ii}}]~$\lambda 6584$}

% Balmer lines
\def\Ha{{H$\alpha$}}
\def\Hb{{H$\beta$}}

% Other lines
\def\HeII{{He\,{\sc ii}}}

% Line ratios
\def\NIIHa{[\mbox{N\,{\sc ii}}]/H$\alpha$}
\def\SIIHa{[\mbox{S\,{\sc ii}}]/H$\alpha$}
\def\OIHa{[\mbox{O\,{\sc i}}]/H$\alpha$}
\def\OIIIHb{[\mbox{O\,{\sc iii}}]/H$\beta$}

% Common terms
\def\Ebmv{E($B-V$)}
\def\LOIII{$L[\mbox{O\,{\sc iii}}]$}
\def\Ledd{${L/L_{\rm Edd}}$}
\def\LOIIIs4{$L[\mbox{O\,{\sc iii}}]$/$\sigma^4$}
\def\LOIIIMbh{$L[\mbox{O\,{\sc iii}}]$/$M_{\rm BH}$}
\def\Mbh{$M_{\rm BH}$}
\def\Msigma{$M_{\rm BH} - \sigma$}
\def\Ms{$M_{\rm *}$}
\def\Msun{$M_{\odot}$}
\def\Msunyr{$M_{\odot}yr^{-1}$}

% Units
\def\ergs{$~\rm erg~s^{-1}$}
\def\kms{$~\rm km~s^{-1}$}

% Software
\def\galfit{\texttt{GALFIT}}
\def\multidrizzle{\texttt{multidrizzle}}

% Other
\def\sersic{S\'{e}rsic}

\date{}

\pagerange{\pageref{firstpage}--\pageref{lastpage}} \pubyear{2016}

\maketitle

\label{firstpage}

\begin{abstract}
In order to perform a more complete census of active galactic nuclei (AGN) in the local Universe, we investigate the use of the \HeII~$\lambda 4685$ emission line diagnostic diagram by \cite{2012MNRAS.421.1043S} in addition to the standard methods based on other optical emission lines. The He II based diagnostics is more sensitive to AGN ionization in the presence of strong star formation than conventional line diagnostics. We survey a magnitude-limited sample of 63,915 galaxies from the Sloan Digital Sky Survey Data Release 7 at $0.02<z<0.05$ and use both the conventional BPT emission line diagnostic diagrams, as well as the \HeII\ diagram to identify AGN. In this sample, 1,075 galaxies are selected as AGN using the BPT diagram, while an additional 234 galaxies are identified as AGN using the \HeII\ diagnostic diagram, representing a 22\% increase of AGN in the parent galaxy sample. We explore the host galaxy properties of these new \HeII\ selected AGN candidates and find that they are most common in star-forming galaxies on the blue cloud and on the main sequence where ionization from star-formation is most likely to mask AGN emission in the BPT lines. We note in particular a high \HeII\ AGN fraction in galaxies above the high-mass end of the main sequence where quenching is expected to occur. We use archival Chandra observations to confirm the AGN nature of candidates selected through \HeII\  based diagnostic.Finally,  we discuss how this technique can help inform galaxy/black hole co-evolution scenarios.
\end{abstract}

\begin{keywords}
galaxies: active;  galaxies: nuclei;  galaxies: Seyfert; galaxies: evolution; galaxies: star formation
\end{keywords}

%-----------------------------------------------------------------------------------------------------------------------------------
\section{Introduction}
\label{sec:intro}

Black holes are recognized as an integral part of galaxy evolution \citep{1988ApJ...325...74S, 1998A&A...331L...1S, 2005ApJ...620L..79S, 2008ApJS..175..356H,2008ApJS..175..390H,{2015MNRAS.450..763M}}. The interaction between black holes and their host galaxies is  highly complex. Many different models providing possible physical explanations have been developed; several recent reviews describe in detail the different approaches explaining the links between the star formation and evolution of galaxies with their black holes \citep[][and references therein]{2012ARA&A..50..455F, 2012AdAst2012E..16T, 2013ARA&A..51..511K, 2014ARA&A..52..589H, 2015A&ARv..23....1B}. Over the last 10 years a large number of publications have addressed possible physical causes \citep{2010ApJ...721..193P, 2012ApJ...757....4P, 2014MNRAS.440..889S}, such as AGN feedback \citep{2007ApJS..173..267S, 2007ApJS..173..357K, 2008MNRAS.385.2049G, 2009ApJ...696..891H, 2015MNRAS.450..763M, 2016A&A...588A..78B}, major mergers \citep{1988ApJ...325...74S, 1996ApJ...464..641M, 2006ApJS..163....1H}, environmental effects \citep{2015MNRAS.448..237W, 2015ApJ...800...24K, 2015Natur.521..192P} and secular processes \citep{2004ARA&A..42..603K, 2011MNRAS.411.2026M, 2013ApJ...779..162C}, for the quenching of star formation in galaxies. All these processes  could explain the bi-modality in colour-mass and colour-magnitude space \citep{2003MNRAS.343..367B, 2004AIPC..743..106B, 2007ApJ...665..265F, 2007ApJS..173..342M, Schawinski:2014aa, 2015MNRAS.446.2144T}.

 Observations of AGN in different wavebands lead to different, sometimes contradictory results. According to current models AGN operate in different modes and the various components of AGN are the source of different types of emissions \citep{1994ApJS...95....1E, 1997ASPC..113.....P, 1999ASPC..164...78K, 2006agna.book.....O, 2013peag.book.....N, 2014ARA&A..52..529Y, 2014ARA&A..52..589H}. Depending on their luminosity and the star formation rate of the host galaxy the emissions from AGN can be contaminated by stellar emissions from the host galaxies. As \cite{2012ApJ...753...30S} pointed out, most surveys for AGN are severely biased towards unobscured (type 1) AGN. The most promising  techniques for a complete census of type 2 AGN  are radio, hard X-ray and infrared selection, but only 10\% of AGN are radio loud \citep{2012ApJ...753...30S}. X-ray observations are less affected by dust absorption and  mid infra red observations allows the detection of both type 1 and type 2 AGN \citep{2005ApJ...631..163S, 2012ApJ...753...30S}. 
Nuclear dust and gas can obscure direct emission from AGN making the selection of obscured and heavily obscured AGN more challenging \citep{1995PASP..107..803U, 2003ApJ...598.1017D,2003ApJ...598.1026D,  2011ApJ...739...57K, 2012AdAst2012E..16T}. The most promising  techniques for a complete census of type 2 AGN  are narrow emission lines, radio, hard X-ray and mid infrared selection, but only 10\% of AGN are radio loud \citep{2012ApJ...753...30S}. X-ray observations are less affected by dust absorption and  mid infrared observations allow the detection of both type 1 and type 2 AGN \citep{2005ApJ...631..163S, 2012ApJ...753...30S}. Narrow emission line diagnostic diagrams are thus a widely used method for selecting Type 2 AGN \citep{2003MNRAS.346.1055K, 2006MNRAS.372..961K, 2007MNRAS.382.1415S, 2014ApJ...788...88J}.

The analysis of galaxy colours can be  extremely difficult  because the light of the galaxies can be contaminated by the dust, gas, and star formation \citep{2008ApJ...683..644S}. At low redshift (z $ < 0.7$) this can be overcome because the central sources in the in galaxies  can be resolved with the \textit{Hubble Space Telescope} or ground based optical images \citep{2009ApJ...692L..19S}. The correct colour determination is critical as the present models are based on a galaxy evolution starting in the star forming blue cloud, progressing through the intermediate zone, the  green valley,  to the inactive galaxies of red sequence. As discussed in \cite{2009ApJ...692L..19S} the different models also constrain the possible lifetimes of AGN and the timing of the quenching of the star formation. In particular, if AGN phases occur solely in the green valley, then they cannot be the cause of quenching as they appear at least several hundred Myr after the quenching event. Finding AGN populations in still blue, star-forming galaxies is therefore vital for testing AGN-driven quenching scenarios \citep{2009ApJ...692L..19S}.

In the present work, we concentrate on the standard emission line methods which are based on the \NIIHa\, and \OIIIHb\ \. ratios \citep{ 1981PASP...93....5B, 1987ApJS...63..295V,2001ApJ...556..121K, Kewley:2006aa, 2013ApJ...774L..10K, 2003MNRAS.346.1055K,  2004ApJS..153....9G, 2004ApJS..153...75G,  2007MNRAS.382.1415S, 2006MNRAS.371..972S, 2008MNRAS.391L..29S, 2014ApJ...788...88J}. We show that using the \HeII\ \. / \Hb\ \. ratio \citep{2012MNRAS.421.1043S} we can increase the number of AGN candidates identified and we analyze this group of additional AGN candidates in detail. We show that this may lead to a better understanding of the role of AGN in quenching the star formation. 

This paper is organized as follows: In section \ref{sec:analysis} we describe under section \ref{sec:sample} the sample selection and then introduce in section \ref{sec:eml}  the two emission line methods used for the analysis; we describe the  standard  BPT diagnostic diagram and the \HeII\, based diagnostic. We combine the two methods and discuss the characteristics of the \HeII\, selected AGN candidates. We discuss in section  \ref{subsubsec:results_HEII} the demographics of the classical standard BPT and the \HeII\, selected AGN and the host galaxy properties in section \ref{sec:hostgals}. In section \ref{subsubsec:luminosity} we analyse the \HeII\ line luminosity. To further confirm the nature of the \HeII\, selected AGN candidates we search for counterparts in the X-rays (section \ref{sec:chan}). In section \ref{sec:discussion} we discuss our findings and their possible implications for the star formation quenching scenarios. We present the summary in section \ref{summary}.

Throughout this paper, we use a standard $\Lambda$CDM Cosmology consistent with observational measurements \citep{2011ApJS..192...18K}. All magnitudes are in the AB system.
%-----------------------------------------------------------------------------------------------------------------------------------

%-----------------------------------------------------------------------------------------------------------------------------------
\section{Analysis}
\label{sec:analysis}

%-----------------------------------------------------------------------------------------------------------------------------------
\subsection{Sample selection}
\label{sec:sample}

We base our sample on the seventh data release (DR7) of the Sloan Digital Sky Survey (SDSS; \citealt{2000AJ....120.1579Y,2009ApJS..182..543A}). We cross match the New York Value-Added Galaxy Catalog (NYU VAGC, \citealt{2005AJ....129.2562B, 2008ApJ...674.1217P}) with the Max Planck Institute for Astrophysics \ John Hopkins University (MPA JHU, \citealt{2003MNRAS.346.1055K, 2004MNRAS.351.1151B}) catalogue to obtain properties such as stellar masses and star formation rates. We used nebular emission line strengths provided by the OSSY catalog \cite{2011ApJS..195...13O}. This yields a total of  605,019 objects. In order to obtain a magnitude limited sample we limit the redshift interval to 0.02 \textless\, z \textless\, 0.05 and only consider the galaxies with absolute petrosian r-band magnitude \textless\, -19.0. The remaining sample consists of  63,915 galaxies. For the BPT diagram we further restrict our sample to the presence of the [OIII],  [NII],  \Ha,\ and  \Hb\ lines with S/N \textgreater\,  3 resulting in a sample of 37,425 galaxies; for the HeII emission line diagnostic diagram the restricted sample based on the presence of the He II, [NII], \Ha,\ and \Hb\ lines  with S/N \textgreater\,  3 is  872 galaxies.

%-----------------------------------------------------------------------------------------------------------------------------------
\subsection{Emission line diagnostic diagrams}
\label{sec:eml}
In this section we use emission line diagnostic diagrams to classify the dominant source of ionization in our galaxy sample, with the goal of performing a more complete census of galaxies hosting AGN. The classic emission-line diagnostic, also known as BPT diagram \citep{1981PASP...93....5B, 1987ApJS...63..295V, 2001ApJ...556..121K, 2003MNRAS.346.1055K} based on the \OIIIHb\ vs \NIIHa\ ratio (with \SIIHa\ and \OIHa\ as variations) is commonly used. The key point of this paper is to apply a newer diagnostic diagram where \OIII\ is replaced by \HeII, since \HeII\ offers a cleaner AGN/star formation separation as proposed and developed by \cite{2012MNRAS.421.1043S}. 

\begin{figure*}
\begin{center}

\includegraphics[width=0.36\textwidth]{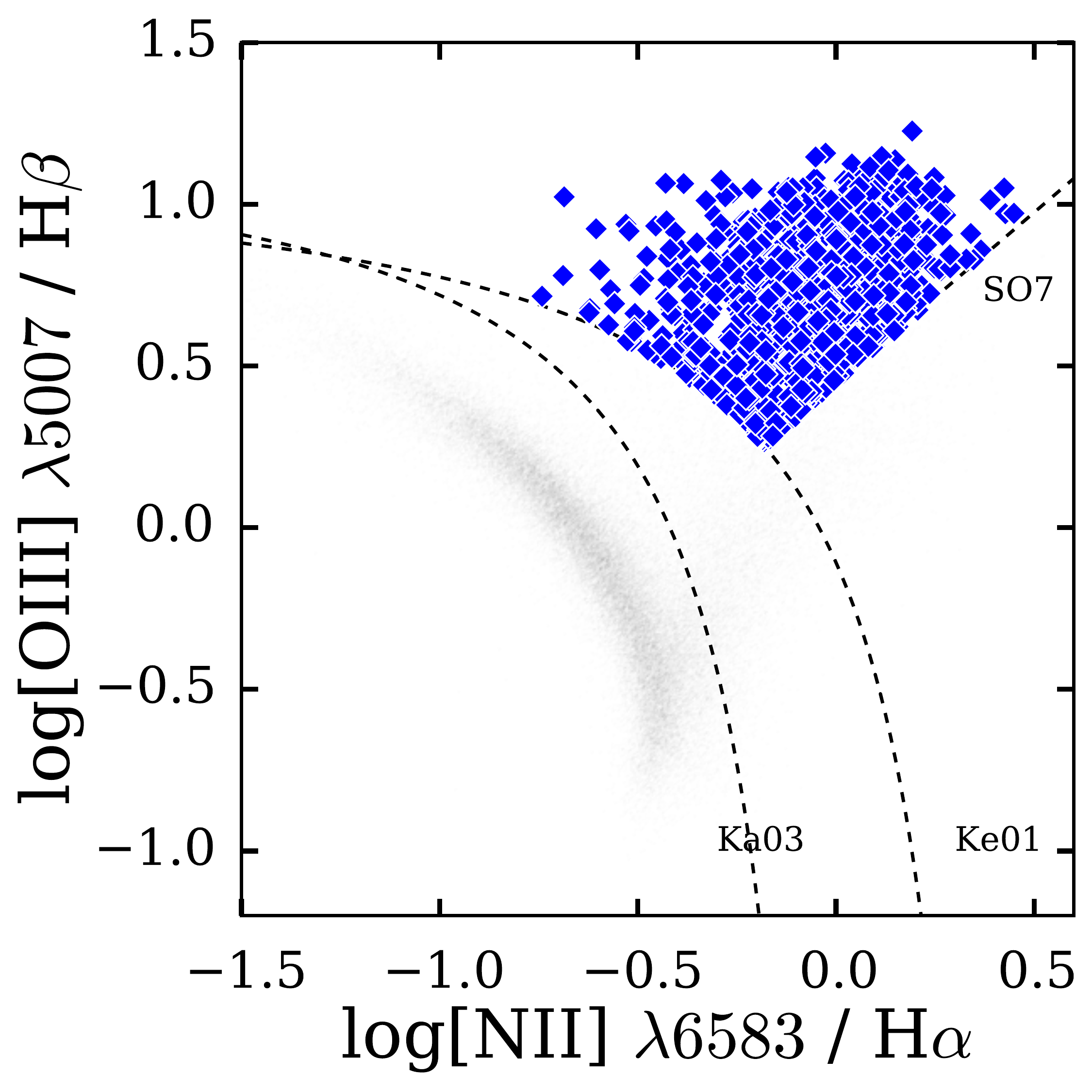}
\includegraphics[width=0.36\textwidth]{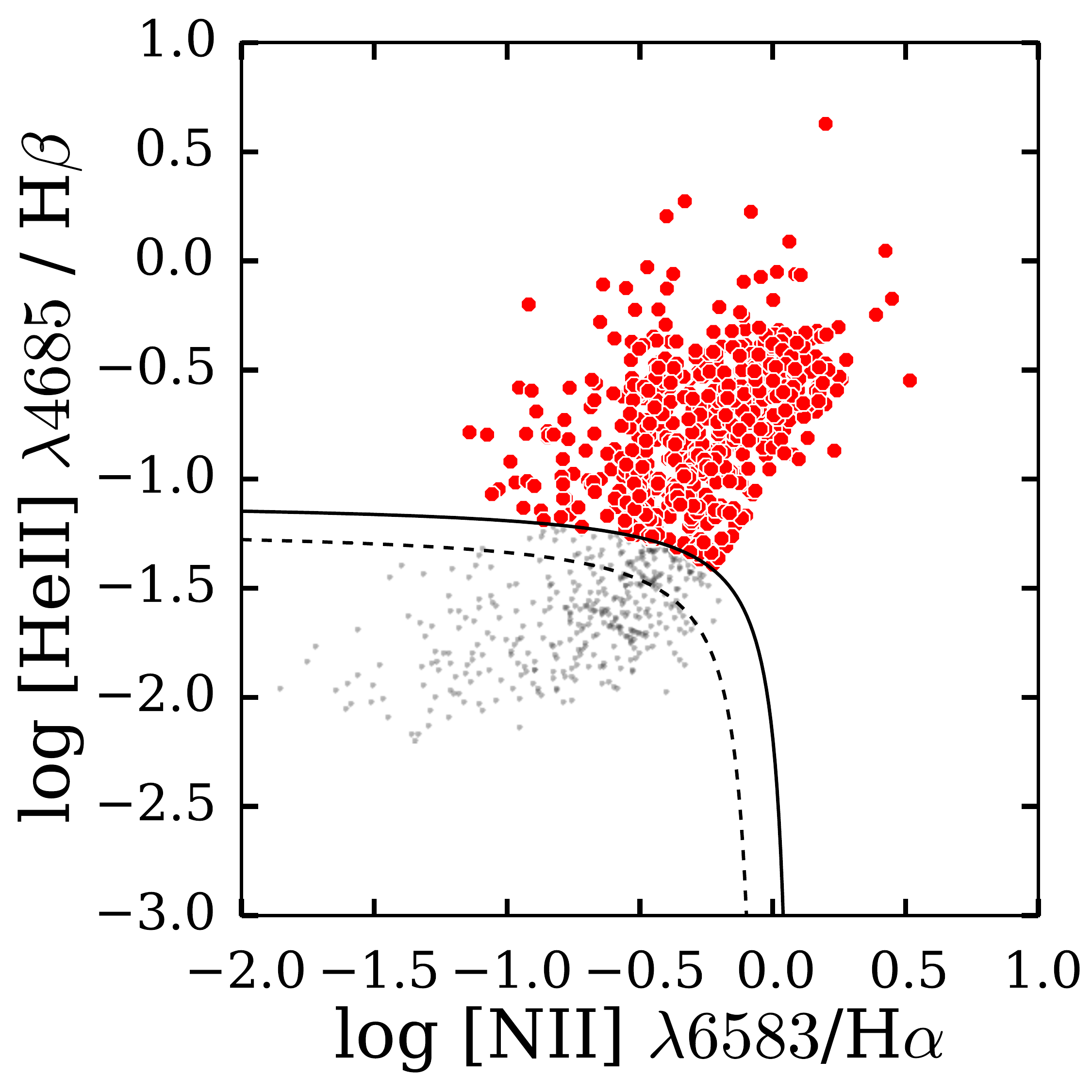}
\includegraphics[width=0.36\textwidth]{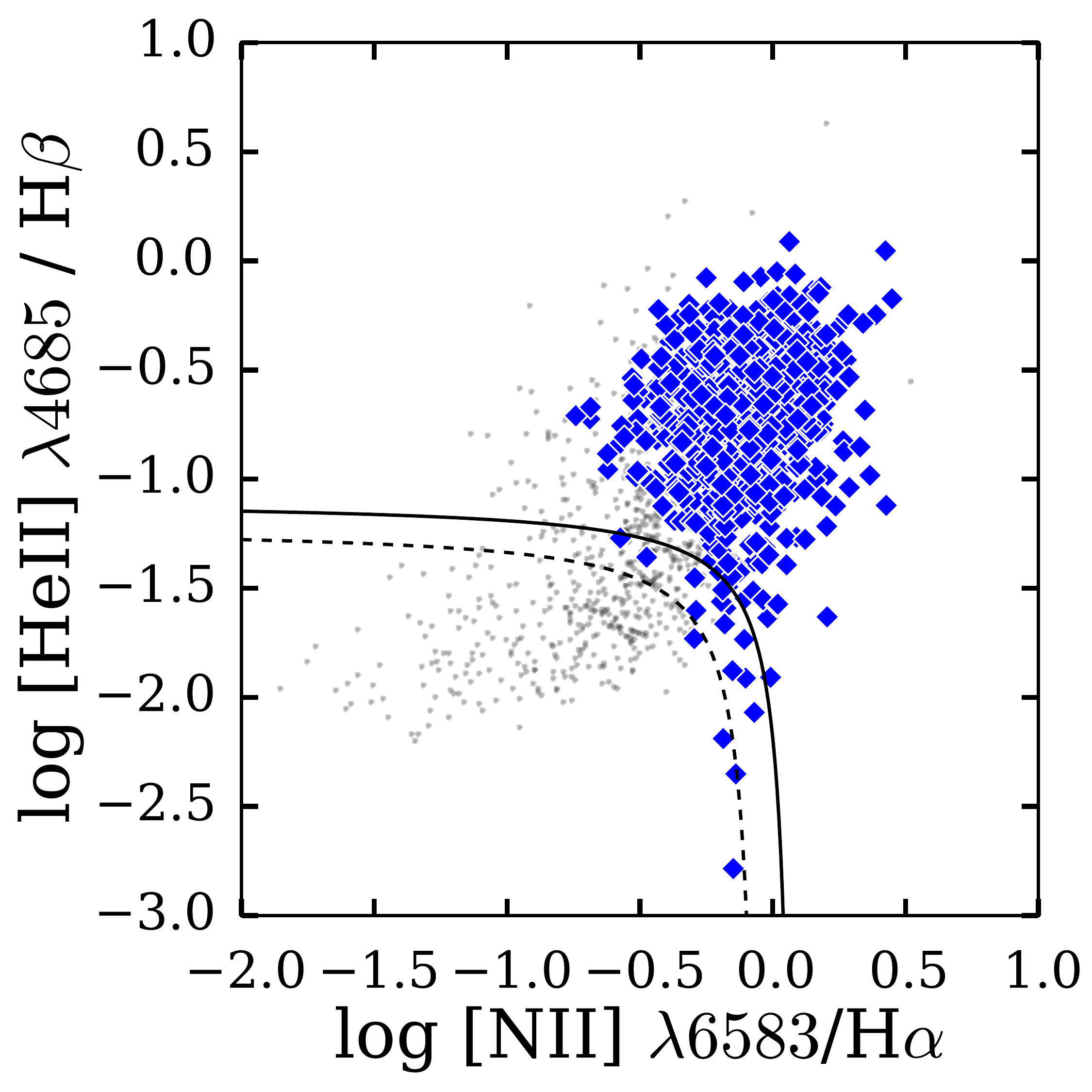}
\includegraphics[width=0.36\textwidth]{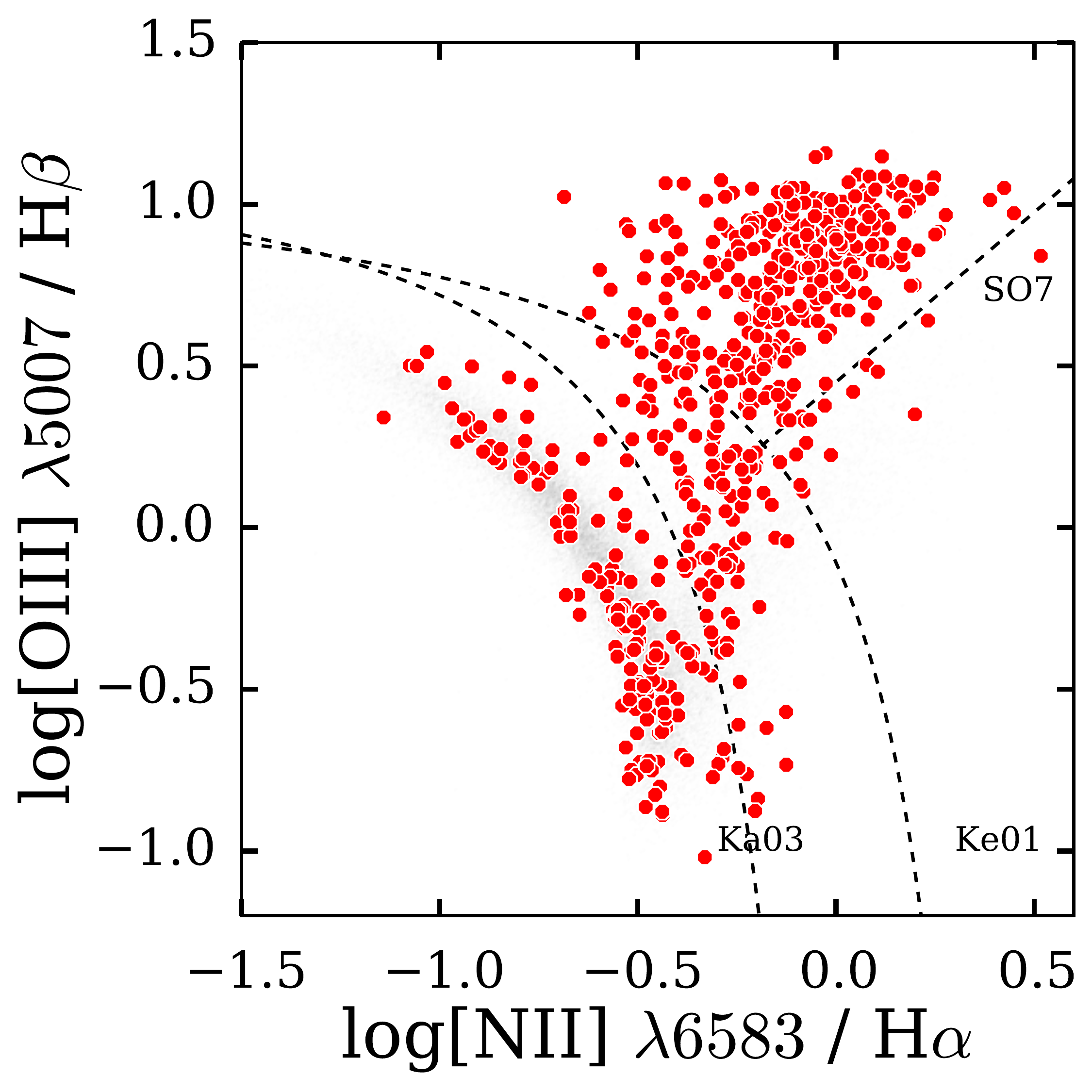}

\caption{The emission line diagnostic diagrams used in this paper. In the \textit{top} row, we show the classic BPT diagram (\OIIIHb\ vs. \NIIHa) (\textit{left}) and the \HeII\ diagram (\textit{right}). AGN candidates selected on the BPT diagram are shown as blue diamonds, while AGN candidates selected on the \HeII\ diagram are shown as red points. The grey shading represents the total  population of AGN candidates.  In the \textit{bottom} row, we show how the respective AGN selections correspond to each other: in the \textit{bottom-left}, we show the \HeII\ diagram with the BPT-selected AGN candidates shown as blue diamonds. 95 \% of the BPT-selected AGN candidates are also selected as AGN candidates using the \HeII\ diagram. In the \textit{bottom-right}, we show the BPT diagram with the \HeII-selected AGN candidates as red points. While some \HeII\ AGN candidates lie in the Seyfert region of the BPT diagram, many do not and scatter across the transition region and into the purely star-forming locus. }

\label{fig:4plot}

\end{center}
\end{figure*}

\subsubsection{Standard BPT diagnostic diagram}
\label{subsubsec:BPT}
As a standard tool, the BPT diagram is widely used to classify galaxies and in particular to identify AGN. It is however not without biases, as the total emission line ratio of a galaxy is driven by the balance of star formation and AGN ionization. This can lead to bias against AGN in highly star-forming galaxies \citep[e.g.][]{2003MNRAS.344L..59M, 2010ApJ...711..284S, 2015MNRAS.454.3722S, 2015ApJ...811...26T,2016ApJ...826...12J}.

We use the standard BPT diagram (\OIIIHb\ vs. \NIIHa)  diagnostic diagram. We identify the population of AGN selected by the standard BPT diagram   as shown in the top left panel of Figure \ref{fig:4plot}; S/N \textgreater\, 3 for the respective emission lines is applied here as well as in all the following diagrams. The grey shading represent our total sample of 63,915 galaxies as defined in section \ref{sec:sample}. The Ke01 line was defined by \cite{2001ApJ...556..121K} as the extreme star burst line; the S07 line as defined by \cite{2007MNRAS.382.1415S}  separates the LINERs from the AGN candidates. We only classify AGN in the Seyfert region of the BPT diagram as BPT AGN, as the majority of LINERs on SDSS BPT diagram are predominantly due to non-AGN processes \citep{2010MNRAS.402.2187S,2012ApJ...747...61Y,2016MNRAS.461.3111B}. This BPT AGN selection from the parent sample yields 1,075 galaxies classified as AGN. 
%-----------------------------------------------------------------------------------------------------------------------------------

%-----------------------------------------------------------------------------------------------------------------------------------
\subsubsection{The \HeII~$\lambda 4685$ emission line diagnostic diagram}
\label{subsubsec:HeII}
\cite{2010ApJ...711..284S} address the weakness of the BPT  diagram in missing low luminosity AGN candidates in star forming galaxies. AGN photoionization can be overwhelmed by star formation, and the ionizing radiation produced my metal poor stars can produce line ratios similar to AGN. The ionizing energy of 54.4 eV for He$^{+}$ is much higher than the ionizing energy of O$^{++}$ (35.2 eV) and N$^{+}$ (15.5 eV) used in the BPT diagram. \HeII\ is also less affected by any gas phase metallicity dependence. This makes it possible to identify AGN candidates using the \HeII\ diagram which cannot be identified using the standard  BPT diagram.  However, the use of the HeII line for diagnostic purposes can be challenging as the HeII line is usually weak.

We therefore use the  \HeII / {{H$\alpha$}} vs {{[\mbox{N\,{\sc ii}}] / {{H$\beta$}} diagnostic diagram following \cite{2012MNRAS.421.1043S}. In the top right panel of Figure \ref{fig:4plot} we show the \HeII\ diagram for our galaxy sample. \cite{2012MNRAS.421.1043S} showed that above the black dotted line more than 10\% of the  \HeII\ flux  and that above the solid black line more than 50 \% of  \HeII\ flux originates from AGN candidates. This solid line is therefore used as the dividing line for the selection of  \HeII\ AGN candidates. Using the \HeII\ diagnostic we selected 559 AGN shown as red points. In the following text will refer to the AGN candidate selected with HeII diagnostic as HeII AGN.
%-----------------------------------------------------------------------------------------------------------------------------------

%-----------------------------------------------------------------------------------------------------------------------------------
\subsubsection{Results of combining the BPT and the Shirazi \& Brinchmann diagnostic}
\label{subsubsec:results}

\begin{table*}
\begin{tabular}{lllllllllll} 

\hline
{SDSS ID}&{RA}&{DEC}&{z}&{MASS}&{COLOUR}&{BPT AGN}&{HeII AGN}&{HeII BPT AGN}&{HeIIonly}&{BAD}\\
\hline
{588848899929014499}&{230.10231}&{-0.2252278}&{0.035}&{9.64}&{1.86}&{1.}&{0.}&{0.}&{0.}&{0.}\\
{587722982835749016}&{217.36247}&{-0.20893236}&{0.028}&{10.12}&{1.69}&{0.}&{1.}&{0.}&{1.}&{0.}\\

\hline

\end{tabular}
\caption{We provide this table of all the BPT selected AGN candidates and the additional AGN candidates selected by the \HeII  method in electronic form. We give the SDSS ID, RA, DEC, z, Mass and Colour for all AGN candidates. The respective categories BPT selected AGN candidates, HeII selected AGN candidates, the AGN candidates selected by both methods and HeII-only AGN candidates have a flag 1. We visually inspected all HeII spectra and set a flag 1 in the "bad" column for those which we considered unreliable.}
\label{Table:3}
\end{table*}

We combine the BPT and \HeII\, method results in the two lower panels of Figure \ref{fig:4plot}. In the lower left panel we show the position of the BPT  AGN candidates on the \HeII\ diagram as blue diamonds. 95\% of the BPT AGN are above the dividing line for AGN candidates selected also as AGN  using the \HeII\ diagnostic.

In Figure \ref{fig:4plot} we also plot in the bottom right panel the \HeII\ AGN candidates as red circles on the standard BPT  [OIII] vs  [NII] diagram.  Their positions extend well into the area below the Ke01 and Ka03 line \citep{2001ApJ...556..121K, 2003MNRAS.346.1055K}  respectively, which puts them into the star forming region. If the \HeII\ AGN candidates are bona-fide AGN, then their location in the star-forming region of the BPT diagram can be explained by strong star formation in the host galaxies overwhelming the AGN signal.

We  provide a complete list of the objects selected by both methods in Table \ref{Table:3}.  
%-----------------------------------------------------------------------------------------------------------------------------------

%-----------------------------------------------------------------------------------------------------------------------------------
\subsection{Demographics of standard BPT and \HeII-selected AGN}
\label{subsubsec:results_HEII}

\begin{figure}
\begin{center}

\includegraphics[width=0.49\textwidth]{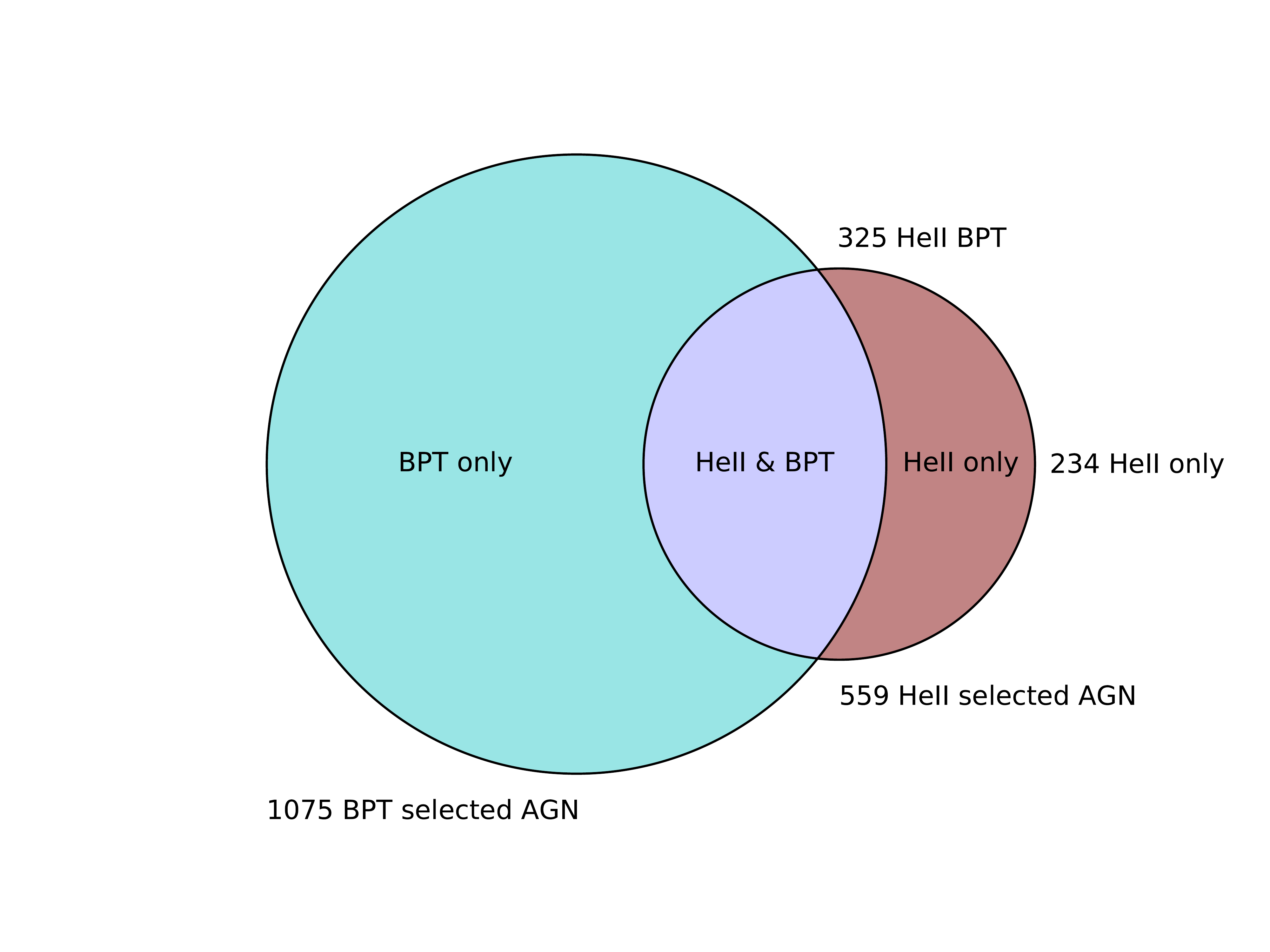}
\caption{With the Venn diagram we show the relationship between the BPT and \HeII\ AGN candidates selections. From our sample of 63'915 galaxies the BPT diagnostic identified 1'075 AGN candidates and the \HeII\ diagnostic, according to \citet{2012MNRAS.421.1043S},  identified 559 AGN candidates. 325 AGN candidates  are found by both methods. The \HeII method identifies 234 AGN candidates with are not detected by the BPT method, which represents an increase of 22\% to the number of AGN candidates identified by the BPT diagram.}

\label{fig:venn}

\end{center}
\end{figure}

In order to summarize the analysis performed in Figure \ref{fig:4plot} we present the results in a Venn diagram in Figure \ref{fig:venn}. We can distinguish four different groups of AGN candidates:

\begin{enumerate}
\item 1,075 galaxies identified as AGN by the standard  BPT diagnostic diagram only (hereafter BPT AGN)
\item 559 galaxies identified as AGN by the HeII diagnostic  diagram (hereafter HeII AGN) 
\item 325  galaxies identified as AGN by both diagrams (hereafter HeII BPT AGN)
\item 234 galaxies identified as AGN candidates by the \HeII\ diagnostic diagram only (hereafter \HeII-only AGN)
\end{enumerate}

As we show here, there exists a group of AGN candidates in the \HeII\ selected sample which the BPT method misses. Before we made the detailed analysis we visually inspected the \HeII\ region of their SDSS spectra and  eliminated 9 spectra where we deemed the fits to the \HeII\ line to be unreliable following \cite{2015ApJS..219....1O}.
 
From the total sample of  63,915 galaxies we identify with the BPT method 1,075 AGN candidates and with the \HeII\ method 559 AGN candidates. Of these 559 AGN 325 overlap with BPT selected AGN candidates; this means that they are selected by both methods. However the \HeII\ method selected 234 \HeII -only AGN candidates not selected by the the BPT} method. This corresponds to an increase in the detection rate of 22 \% compared to the 1,075 AGN candidates identified by the BPT method (see Figure \ref{fig:venn}).
%-----------------------------------------------------------------------------------------------------------------------------------

%-----------------------------------------------------------------------------------------------------------------------------------
\subsection{Host galaxy properties}
\label{sec:hostgals}

\begin{figure*}
\begin{center}

\includegraphics[width=\textwidth]{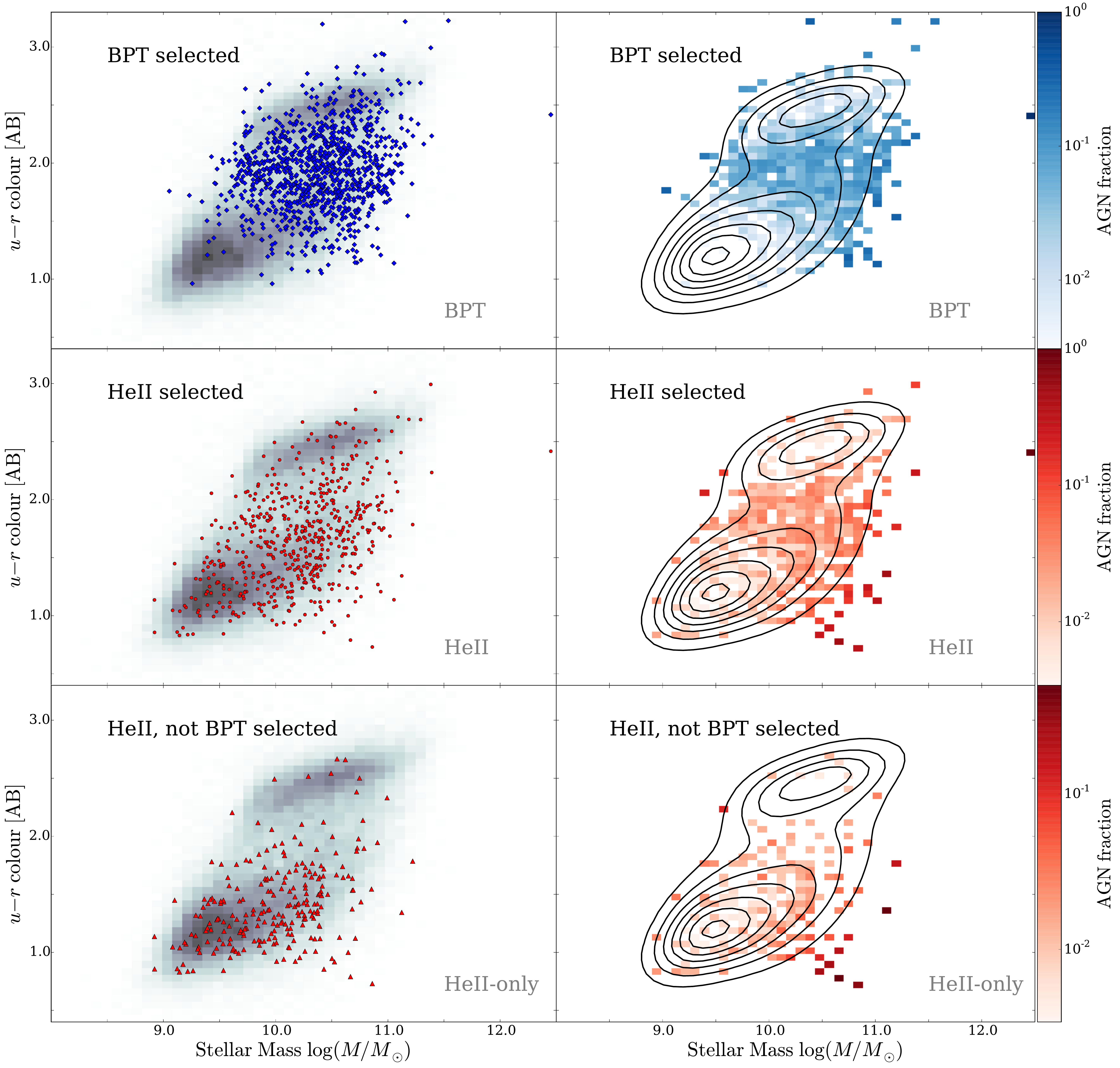}
\caption{We show galaxy colour-mass diagrams with the various AGN selections shown in the different panel. In the top panel of the left column, the BPT AGN candidates are shown as blue diamonds. They are predominantly located in the green valley. In the middle panel of the left column, the \HeII\ selected AGN candidates are shown as red points. They are predominantly located in the green valley and scatter into the blue cloud. In the bottom panel of the left column we show those  \HeII\  selected AGN candidates \textit{which are not identified by the BPT method} as red triangles. They are concentrated in the blue cloud. The \HeII\ \textcolor{red}{-} only AGN candidates host galaxies are less massive than those selected by BPT diagram. In the right hand column we show the respective fractions of AGN from the various selection methods. These AGN fraction colour-mass diagrams highlight the concentration of AGN hosts in the green valley, with the\HeII\-only AGN being the only population with an elevated fraction in the blue cloud. The fraction plots also highlight that even the prevalence of \HeII\ - only AGN in the blue cloud do not lead to an elevated overall AGN fraction in highly star-forming galaxies -- there are too many underlying star-forming galaxies.}

\label{fig:6plot}

\end{center}
\end{figure*}

\begin{figure*}
\begin{center}

\includegraphics[width=0.49\textwidth]{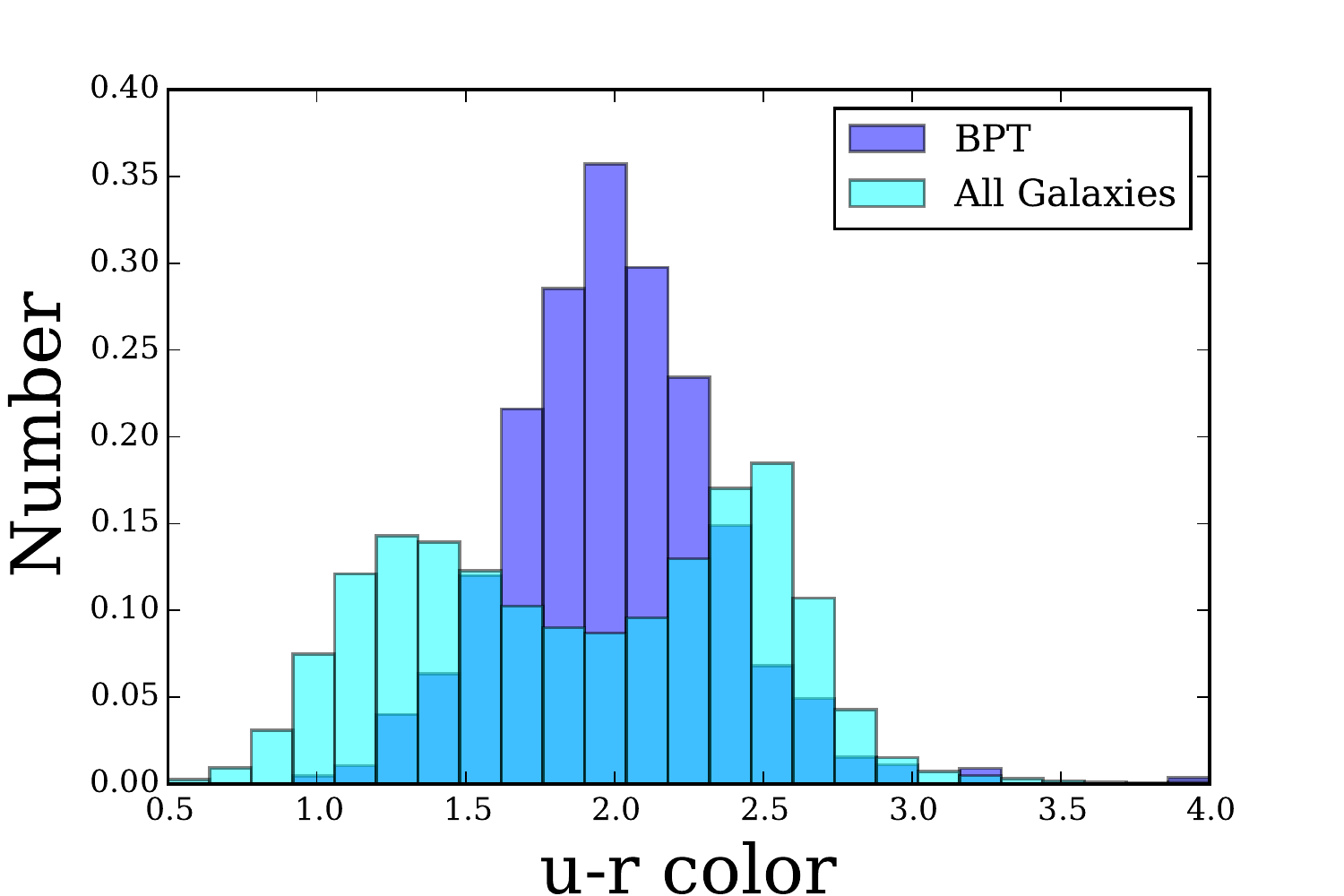}
\includegraphics[width=0.49\textwidth]{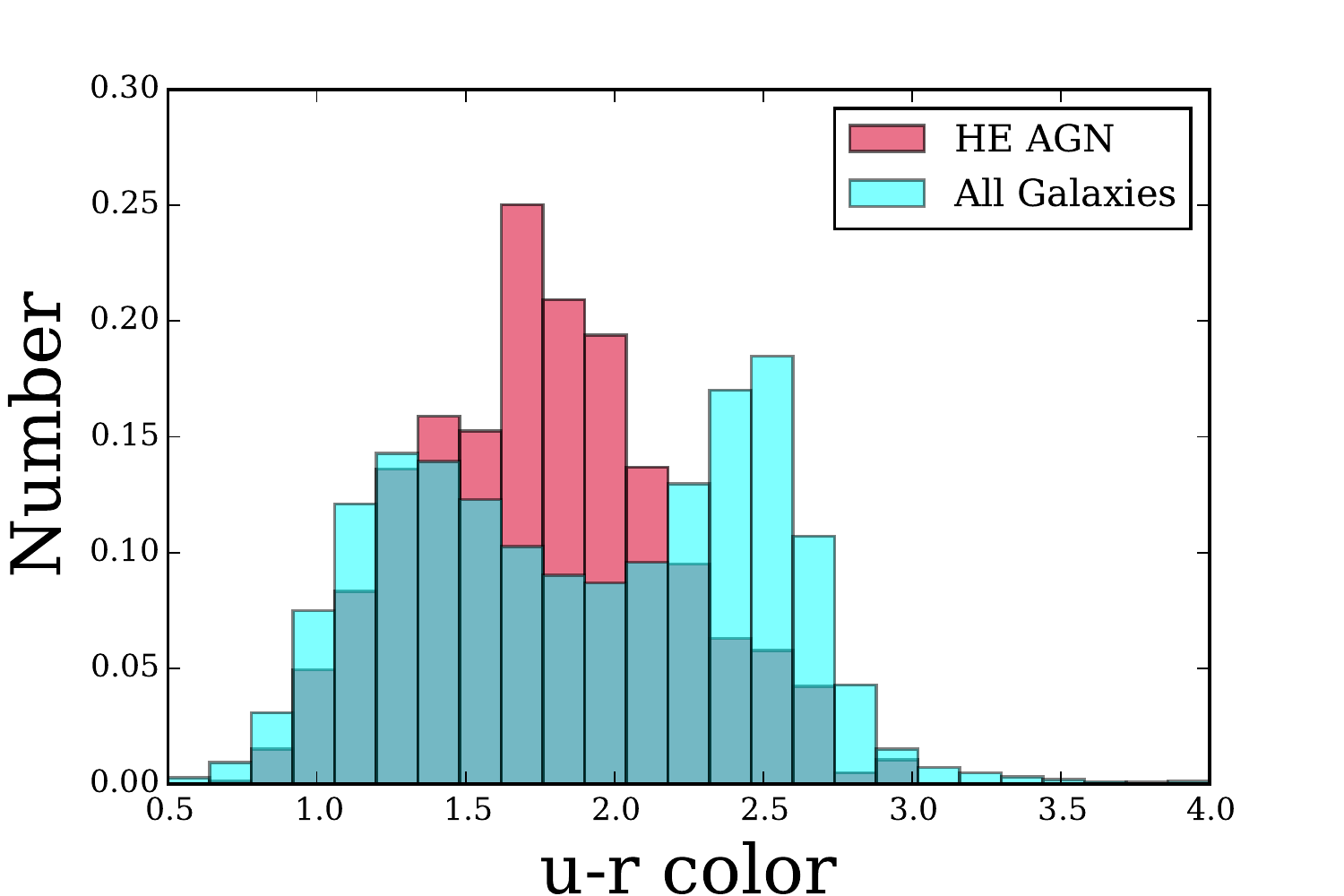}
\caption{We show the $u-r$ colour histograms of the BPT selected and \HeII\ AGN host galaxies, and compare them to the total galaxy sample. The \HeII\ selected AGN candidates are bluer than the BPT selected AGN host galaxies. }

\label{fig:Hist}

\end{center}
\end{figure*}

\begin{figure}
\begin{center}

\includegraphics[width=0.49\textwidth]{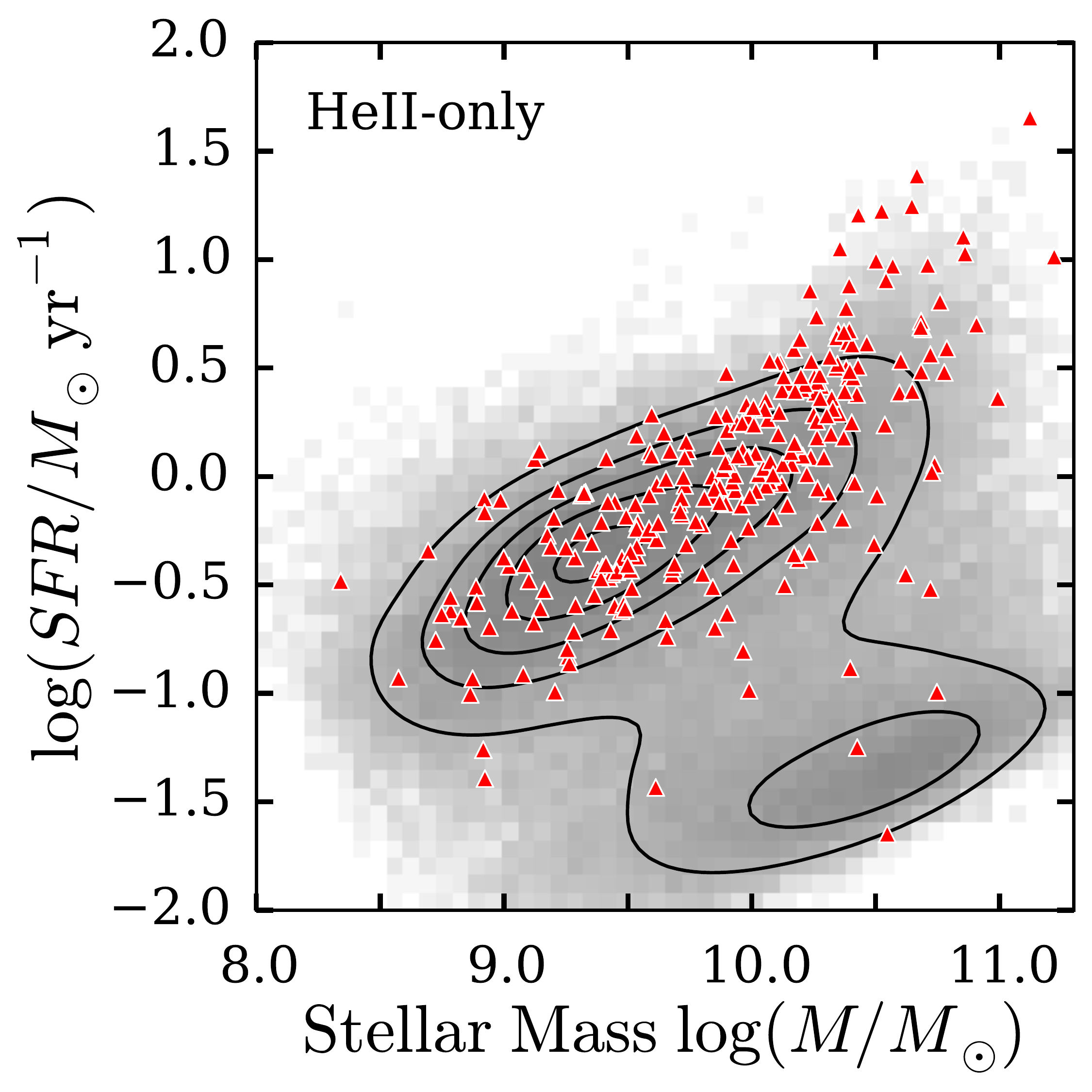}
\caption{We show that the \HeII-only AGN which are \textit{not} selected on the standard BPT diagram reside predominantly on the main sequence. The \HeII-only AGN hosts show increased star formation rates above the main sequence for stellar masses above $10^{10.5}$\Msun. According to \protect\cite{2011ApJ...739L..40R} star bursts in high mass galaxies are thought to be merger driven; they may represent a critical phase towards the quenching of star formation and morphological transformation in galaxies.  They grey-shaded contours show galaxies classified as star-forming according to the BPT emission line diagram.} 

\label{fig:msfr}

\end{center}
\end{figure}

\begin{figure}
\begin{center}

\includegraphics[width=0.48\textwidth]{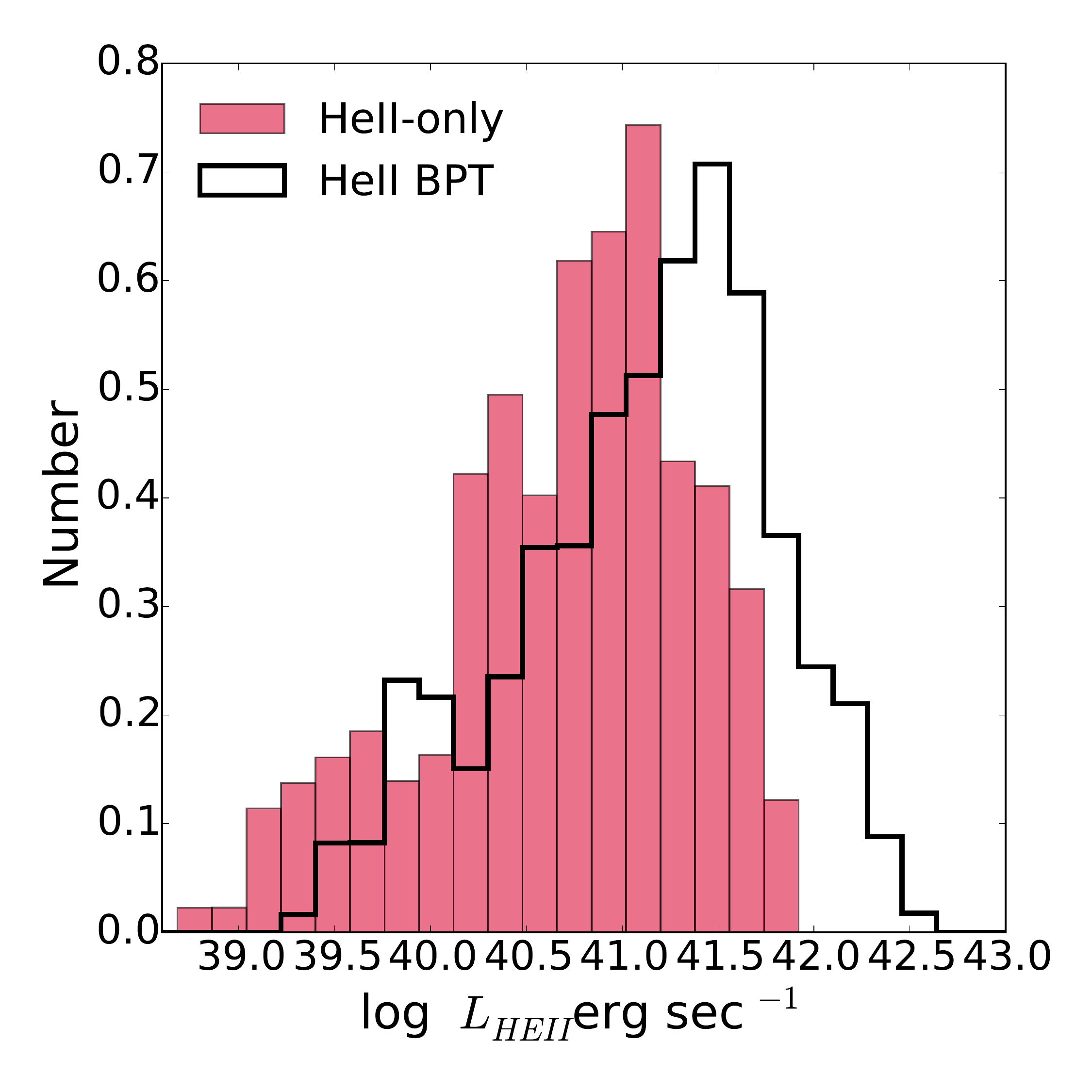}
\caption{We plot the histogram of the \HeII\ line luminosity for the \HeII-only AGN and for the objects which are selected by the standard BPT and the \HeII\ selection. The mean luminosity of the \HeII-only AGN is 40.71\ergs\ and 41.10 \ergs\ for the AGN detected by both. Taking the \HeII\ line luminosity as a proxy for AGN luminosity \citep{2015MNRAS.454.3622B}, we find that the \HeII\ only objects are $\sim0.4$ dex less luminous than the objects detected by both methods.}

\label{fig:Hist_LU}

\end{center}
\end{figure}

To analyze the characteristics of the different groups of AGN candidates we plot the galaxy colour-mass diagram in Figure \ref{fig:6plot}. We show both the distribution of AGN host galaxies from the two selection methods and the implied AGN fraction as a function of colour and mass. We recover the well-known result that AGN host galaxies tend to lie in the optical green valley  both in terms of distribution, and more strongly in terms of AGN fraction.  \citep{2007ApJ...660L..11N, 2007ApJS..173..267S, 2007MNRAS.382.1415S,  2009ApJ...692L..19S, 2010ApJ...711..284S, 2014MNRAS.440..889S, 2008ApJ...675.1025S, 2009ApJ...696..891H}.

Both the standard BPT selection and the \HeII\ AGN exhibit this behaviour. When we consider the \HeII-only AGN however (the red triangles), we find that they preferentially reside in star-forming galaxies in the blue cloud and largely avoid the green valley. The \HeII-only AGN fraction in the blue cloud reaches 10$^{-1}$ but does not reach the AGN fraction seen in the green valley using other methods -- this is due to the large numbers of underlying star-forming galaxies in the blue cloud. We verify this behaviour by plotting the $u-r$ colour histograms of the BPT and \HeII\ AGN in Figure \ref{fig:Hist}.

This observation can be further explored by plotting the \HeII-only AGN hosts on the ``main sequence" diagram of stellar mass versus star formation rate \citep{2007ApJ...660L..47N, 2011A&A...533A.119E, 2015ApJ...801...80L}. We show the main sequence in Figure \ref{fig:msfr} with the \HeII-only AGN hosts as red triangles. As these objects are classified as star-forming on the standard BPT diagram, we can interpret their \Ha-derived star formation rates as reliable. We note that above $M^{*}$ (log $M^{*}$ $\sim$ 10.8 e.g. \citealt{2016MNRAS.459.2150W}), the \HeII-only AGN hosts are not just on the main sequence, but systematically elevated above it.

All this points to a picture where the \HeII\ AGN selection is more sensitive to AGN activity in highly star-forming galaxies than the standard BPT method, and the \HeII-only AGN hosts missed by the BPT method are highly star-forming galaxies in the blue cloud and on the main sequence, with the most massive \HeII-only AGN hosts being significantly elevated from the main sequence.
%-----------------------------------------------------------------------------------------------------------------------------------

%-----------------------------------------------------------------------------------------------------------------------------------
\subsection{Analysis of \HeII\ line luminosity }
\label{subsubsec:luminosity}

\cite{2015MNRAS.454.3622B} have shown that the \HeII\ line luminosity correlates with the AGN hard X-ray luminosity  as well as other lines commonly used \citep{1987ApJS...63..295V,2010MNRAS.402.2187S,2010ApJ...711..284S}, such as [OIII]. We can therefore explore whether the \HeII-only AGN are more or less luminous than the \HeII\ AGN which are also detected by the standard BPT method. We plot the histogram of \HeII\ line luminosities in Figure  {\ref{fig:Hist_LU} for both these populations. We find that the mean luminosity of the \HeII-only AGN is 40.71\ergs\ and 41.10 \ergs\ for the AGN detected by both. This means that the \HeII-only AGN are intrinsically less luminous, even though they are hosted by more highly star-forming galaxies. If they were more luminous, they likely would also be detected by the BPT method. This does however not explain why the host galaxies are bluer: this could due to the BPT selection bias against low luminosity AGN in star forming galaxies; another explanation could be that  the black hole masses of blue cloud AGN hosts are generally lower.
%-----------------------------------------------------------------------------------------------------------------------------------

%-----------------------------------------------------------------------------------------------------------------------------------
\subsection{Analysis of Archival \textit{Chandra} X-ray Observations}
\label{sec:chan}

\begin{figure*}
\begin{center}

\includegraphics[width=0.49\textwidth]{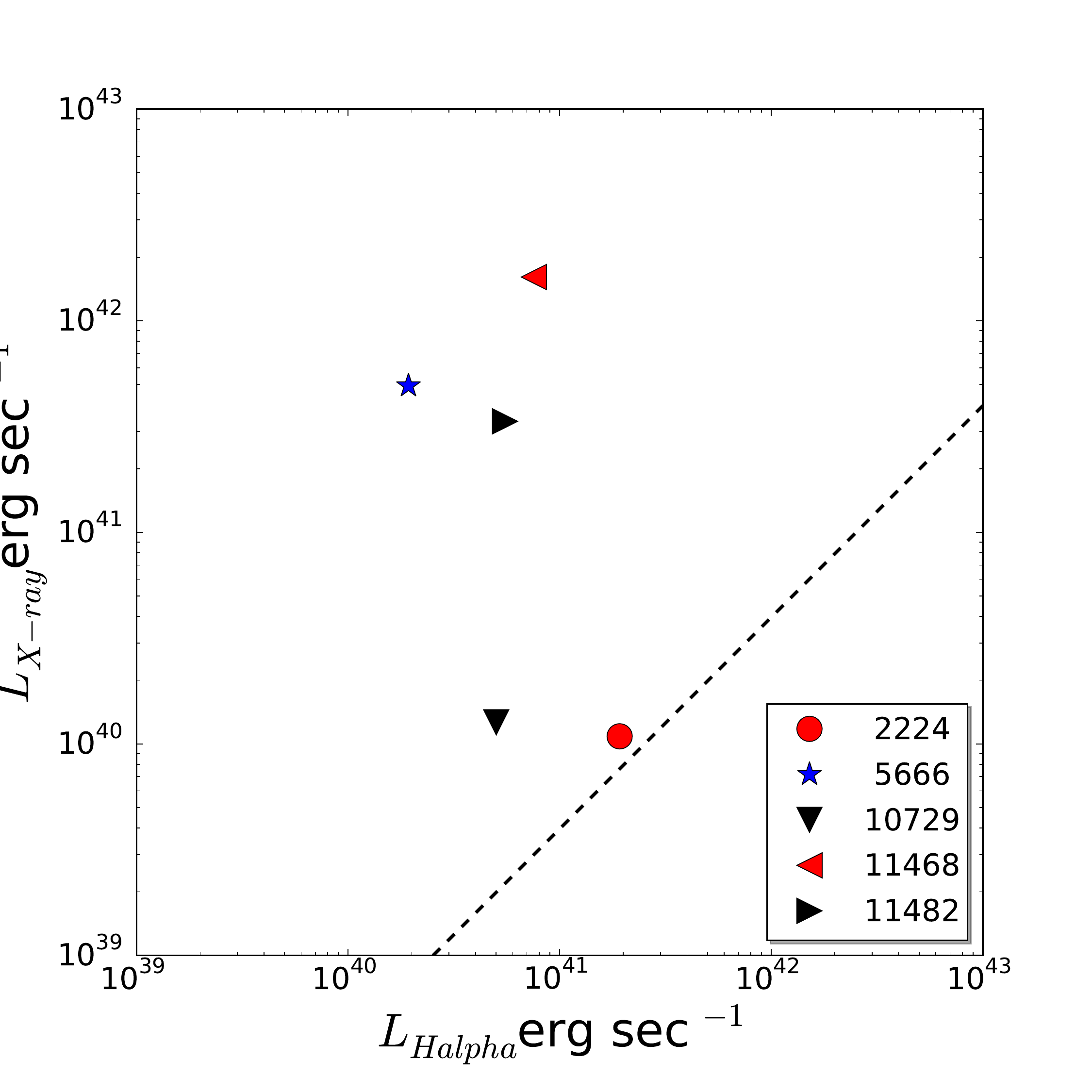}
\includegraphics[width=0.49\textwidth]{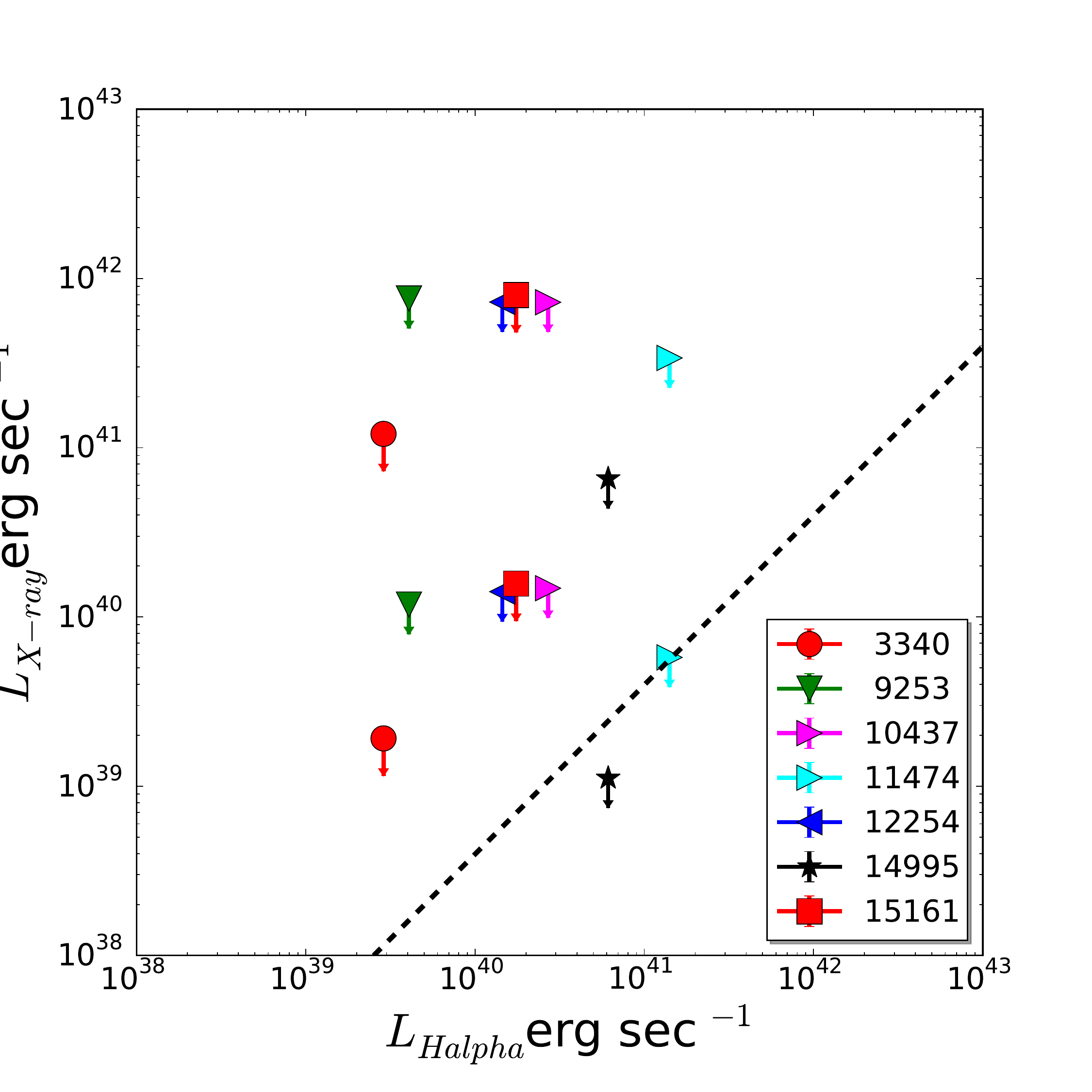}

\caption{We compare the observed X-ray luminosity of those \HeII-only AGN with archival observations and compare them to the observed H$\alpha$ luminosity to test whether the observed X-ray emission could come from star formation alone. In the \textit{left} panel, we show the 5 \HeII-only AGN for which we could obtain X-ray spectra. The dashes line shows the expected level of X-ray emission given the H$\alpha$ luminosity from star formation alone from \citet{2003A&A...399...39R}. The plot shows that the X-ray derived luminosity for the 5 objects are above the dashed line, so star formation as a source for the X-ray emission can be ruled out.  In the \textit{right} panel, we show the X-ray luminosity upper limits for 7 other \HeII-only AGN with X-ray coverage. For each object the upper limit for the 2 models with Intrinsic N\textsubscript{H} of Galactic plus zero and of Galactic plus {$10^{23} $} $cm^{-2}$ is shown. 6 objects have the upper limits in both cases above the dashed line, 1 object has the upper limit for the higher N\textsubscript{H} value below the dashed line.  Thus 6 out of 7 objects are consistent with their X-ray emission coming from an AGN rather than star formation. The Observation IDs  are shown next to the symbols in the legends.}  

\label{fig:HRvsHa}

\end{center}
\end{figure*}

\begin{table*}
\begin{tabular}{llllllll}
\hline
{Observation ID} & {Exposure} & {Counts} & {Model} & {red $\chi^2$} & {Flux 2--10 keV} & {Luminosity 2--10 keV} & {NH/100} \\
{} & {ksec} & {} & {} & {} & {$\rm erg~s^{-1}~cm^{2}$} & {$\rm erg~s^{-1}$} & {$\rm 10^{20}\,  cm^{-2}$ } \\

\hline
{2224} & {29.77} & {302} & {\texttt{phabs*zpow}} & {1.239} & {$6.215\times 10^{-15}$} & {$1,921\times 10^{41} $} & {0.0421}\\
{5666} & {4.65} & {308} & {\texttt{phabs*zpw}} & {0.9808} & {$2.676\times 10^{-13}$} & {$1.2929\times 10^{40} $} & {0.0259}\\
{10729} & {9.52} & {112} & {\texttt{phabs*diskbb}} & {1.690} & {$3.085\times 10^{-15}$} & {$5.015\times 10^{40} $} & {0.0159}\\
{11468} & {1.98} & {314} & {\texttt{phabs*zpow}} & {0.867} & {$2.858\times 10^{-13}$} & {$7.552\times 10^{40} $} & {0.0448}\\
{11482} & {1.95} & {38?} & {\texttt{phabs*zpow}} & {1.05} & {$1.241\times 10^{-13}$} & {$5.519\times 10^{40} $} & {0.0135}\\

\hline
\end{tabular}
\caption{In this table we provide the information for the 5 \HeII-only AGN for which we found archival $Chandra$ observations and which had sufficient counts to extract spectra. The dashed line corresponds to the expected X-ray luminosity assuming that all the H$\alpha$ luminosity comes from star formation.}
\label{Table:1}
\end{table*}

\begin{table*}
\begin{tabular}{lll}
\hline
{Observation ID} & {F[2-10]keV gal+none} & {F[2-10]keV {gal + $10^{23}$}}\\ 
{} & {[erg/s/cm$^2$]} & {[erg/s/cm$^2$]} \\
\hline
{3340} & {$8.831\times 10^{-16}$} & {$5.5600\times 10^{-14}$} \\
{9253} &  {$7.478\times 10^{-15}$} & {$4.805\times 10^{-13}$} \\
{10437} & {$4.9920\times 10^{-15}$} & {$2.4390\times 10^{-13}$} \\
{11474} & {$1.244\times 10^{-15}$} & {$7.3110\times 10^{-14}$} \\
{12254} & {$9.9597\times 10^{-15}$} & {$4.933\times 10^{-13}$} \\
{14995} &  {$2.3969\times 10^{-16}$} & {$1.407\times 10^{-14}$} \\
{15161} &  {$8.573\times 10^{-15}$} & {$4.344\times 10^{-13}$} \\

\hline
\end{tabular}
\caption{For the 7 \HeII-only AGN for which we have no detection in archival $Chandra$ observations, we provide the X-ray flux in the 2--10 keV range for 2 alternative Intrinsic  N\textsubscript{H} values, Galactic plus none and Galactic plus $10^{23}$ $cm^{-2}$.}
\label{Table:2}
\end{table*}

Optical emission line diagnostics have to be supplemented with observations at other wavelengths to confirm AGN activity of our \HeII-only AGN candidates. In order to obtain archival X-ray observations for our sources we cross-matched the sample of 234 \HeII-only AGN candidates with the \textit{Chandra} Source Catalog \citep{2010ApJS..189...37E}. 
We visually crossmatched the optical positions with the X-ray images using a 2 arcsec search radius and  obtained the X-ray observations for 12 \HeII-only AGN candidates.

The goal of this analysis was to determine the X-ray flux of the AGN candidates and, if possible, their spectral shape. First, the data were reduced with the CIAO 4.7 software (Fruscione et al. 2006, SPIE, 6270) following standard procedure. Since AGN are considered to be point sources in the X-ray but the \texttt{Chandra} PSF varies with off-axis angle, we then determined the extraction aperture to be considered for each source according to their off-axis angle and the Encircled Energy Average Radius reported in the Chandra Proposer's Observatory Guide. For the background estimation we considered three source-free circular regions with a radius of 15 arcsec.

Among 12 sources with  X-ray data available, 5 were detected with sufficient counts to allow us to extract a spectrum and fit it. We extracted the spectra with the CIAO tool \texttt{SPECEXTRACT} and binned them with a minimum of 3 cts/ bin. We then fitted the  obtained spectra with the HEARSAC package \texttt{Xspec} \citep{1996ASPC..101...17A} using Cash statistics \citep{1986ApJ...303..336G}. All 5 spectra show a clear power law with photon indices between 1.62 and 2.11. For each object we determined the intrinsic flux in the 2.0 - 10.0 keV range using the \texttt{zpow }model of \texttt{Xspec}  and derived the respective X-ray luminosity. We show the results in table \ref{Table:1}. We provide the \textit{Chandra} Observation ID, the exposure time the number of counts and the model which we used to fit the spectrum. In addition we show the $\chi_{red}^2$ and the Flux$_{ 2.0-10}$  obtained for the respective model, the resulting luminosity and the absorbing column N$_{\rm H}$. 

While the power law X-ray spectra are indicative of AGN emission, the observed X-ray luminosities in the range of $10^{40-41}$\ergs\ are low for AGN and could potentially be due to star formation. Following \cite{1998ApJ...498..541K}, we therefore computed what the expected X-ray emission from star formation \textit{would be} if all the intrinsic X-ray and H$\alpha$ luminosity is due to star formation instead of AGN activity (\citealt{2003A&A...399...39R}, \citealt{1998ApJ...498..541K}). In the left panel of Figure \ref {fig:HRvsHa}, we plot the intrinsic 2.0 - 10.0 keV Luminosity versus H$\alpha$ luminosity for the 5 \HeII-only AGN with X-ray spectra and find that the X-ray luminosity of three objects (\textit{Chandra} OBS ID 11468, 5666 and 11482) exceed the expected emission from star formation by more than 2 orders of magnitude; the other 2 objects (\textit{Chandra }OBS ID  2224 and 10729) are also above the line, but their emission may still be consistent with star formation.
For 7 further \HeII-only AGN for which we detected no emission we determined upper limits for the 2.0--10.0 keV flux. First we estimated the noise level (in counts per second) in the source region from the measured background counts and computed the 3$\sigma$ upper limits as 3 times the noise level. We then converted the obtained counts rate to flux with the WebPIMMS\footnote{\url{https://heasarc.gsfc.nasa.gov/cgi-bin/Tools/w3pimms/w3pimms.pl}} simulator assuming a power law model with a photon index  $\Gamma$ = 1.9. For this calculation we determined the Galactic N\textsubscript{H} using the \textit{Colden Galactic Neutral  Hydrogen Calculator} and applied 2 different values for the intrinsic N\textsubscript{H}:  N\textsubscript{H intr} = 0 and N\textsubscript{H intr} = {$ 10^{23} $} $cm^{-2}$. The results for these 7 objects are given in table \ref{Table:2}, where we list the \textit{Chandra} Observation ID, and the $F_{2-10 KeV}$  for both alternatives given above. 

Similar to what we did for the sources with detected X-ray emission, we compared the observed X-ray luminosity upper limits to the X-ray emission from star formation alone in the right panel of Figure \ref {fig:HRvsHa}. For the model with  N\textsubscript{H intr} =  {$10^{23} $} $cm^{-2}$ the upper limits are 2 to 4 orders of magnitude above the dashed line and so these non-detected \HeII-only AGN candidates are consistent with being AGN. 
%-----------------------------------------------------------------------------------------------------------------------------------

%-----------------------------------------------------------------------------------------------------------------------------------
\section{Discussion}
\label{sec:discussion}
In this paper, we have applied the \cite{2012MNRAS.421.1043S} \HeII\ diagnostic diagram to the local Universe galaxy population to perform a more complete census of black hole growth in galaxies, particularly in highly star-forming galaxies. This analysis should be seen as an addition to the ongoing discussion about galaxy evolution scenarios, in particular to star formation quenching mechanisms. 

The analysis of \HeII-only AGN host galaxies compared to AGN host galaxies from the standard} BPT selection (Figure \ref{fig:6plot}) shows that the host galaxies of AGN missed by the BPT selection and detected only by the \HeII\ method are situated primarily in the blue cloud in the colour-mass diagram. This is equivalent to the \HeII-only AGN residing primarily on the main sequence of star formation (Figure \ref{fig:msfr}) The position of AGN candidates in colour-mass space is important for understanding their role in star formation quenching.  

If AGN reside exclusively in green valley galaxies, then it would imply that they are not relevant for the quenching process since their host galaxies have already experienced the quenching "event" several hundred million years in the past \citep{2007MNRAS.382.1415S, 2009ApJ...692L..19S, 2014MNRAS.440..889S, 2007MNRAS.381..543W, 2010MNRAS.405..933W}. However, if there are AGN in galaxies which are still blue and star-forming, then they are plausible sites for AGN-driven quenching.

As we have discussed above, previous results based on the BPT selection find that BPT selection is biased against AGN in blue and star forming galaxies \citep{2015ApJ...811...26T, 2015MNRAS.454.3722S}. Therefore finding AGN mainly in the green valley might only be the result of detecting AGN preferentially in host galaxies in which the star formation has dropped enough to allow the detection of the AGN. The \HeII-selected AGN are significantly less affected by this effect, and so the \HeII-only AGN reside preferentially in lower mass, blue cloud galaxies on or above the main sequence. In fact, at the high mass end of the main sequence above $M^{*}$, the \HeII-only AGN are hosted by galaxies which are \textit{above} the main sequence; these galaxies are prime sites to look for evidence of AGN-driven outflows and quenching.

The \HeII\ diagnostic diagram is shown to be a useful tool for AGN selection; it highlights a population of AGN which were previously difficult or impossible to identify because they are located in starforming galaxies. In surveys with high quality spectroscopic data, the \HeII can be measured and used for AGN selection, though the weakness of the \HeII\ line makes it less useful in low signal-to-noise data and at higher redshift.
%-----------------------------------------------------------------------------------------------------------------------------------

%-----------------------------------------------------------------------------------------------------------------------------------
\section{Summary}
\label{summary}
We used the \HeII\ diagnostic diagram developed by \cite{2012MNRAS.421.1043S} to search nuclear activity in the galaxy population of the local universe. We find:

\begin{itemize}
\item In a sample of 63,915 SDSS galaxies, we find 1,075 AGN using the standard BPT selection whereas we find with the HeII diagnostic diagram 559 AGN; of these 234 AGN are only identified by the HeII method (HeII-only), representing an increase of 22\% in the AGN population.
\item We investigate archival Chandra X-ray observations for 12 He II selected AGN candidates”, 5 objects (42\%) are confirmed as AGN based on their X-ray luminosity and power law nature; of the remaining 7 objects, 6 objects have X-ray luminosity upper limits consistent with being AGN. This small sample is of course not representative of the whole HeII-only population.
\item The host galaxies of the \HeII-only AGN are systematically bluer than those selected by the BPT method and they reside in the blue cloud rather than green valley host galaxies.
\item The \HeII-only host galaxies lie on the main sequence of star formation and scatter above it at host galaxy masses above  $M^{*}$.
\item The fact that the host galaxies of the \HeII-only AGN are highly star-forming makes them prime sites for searching for evidence of AGN quenching.
\item The \HeII\ AGN selection method clearly yields valuable insights, but can be challenging to apply to low quality spectroscopic data due to the \HeII\ line being weak.
\end{itemize}

%-----------------------------------------------------------------------------------------------------------------------------------

%-----------------------------------------------------------------------------------------------------------------------------------
\section*{Acknowledgements}
We thank the anonymous referee for helpful comments. LFS, AKW, and KS gratefully acknowledge support from Swiss National Science Foundation Grant PP00P2\_138979/1.  MK gratefully acknowledges support from Ambizione fellowship grant PZ00P2\textunderscore154799/1.   This research has made use of the NASA/IPAC Extragalactic Database (NED) which is operated by the Jet Propulsion Laboratory, California Institute of Technology, under contract with the National Aeronautics and Space Administration.  This research has made use of data obtained from the Chandra Data Archive and the Chandra Source Catalog, and software provided by the Chandra X-ray Center (CXC) in the application packages CIAO, ChIPS, and Sherpa.  Funding for SDSS-III has been provided by the Alfred P. Sloan Foundation, the Participating Institutions, the National Science Foundation, and the U.S. Department of Energy Office of Science. The SDSS-III web site is http://www.sdss3.org/.  SDSS-III is managed by the Astrophysical Research Consortium for the Participating Institutions of the SDSS-III Collaboration including the University of Arizona, the Brazilian Participation Group, Brookhaven National Laboratory, University of Cambridge, Carnegie Mellon University, University of Florida, the French Participation Group, the German Participation Group, Harvard University, the Instituto de Astrofisica de Canarias, the Michigan State/Notre Dame/JINA Participation Group, Johns Hopkins University, Lawrence Berkeley National Laboratory, Max Planck Institute for Astrophysics, Max Planck Institute for Extraterrestrial Physics, New Mexico State University, New York University, Ohio State University, Pennsylvania State University, University of Portsmouth, Princeton University, the Spanish Participation Group, University of Tokyo, University of Utah, Vanderbilt University, University of Virginia, University of Washington, and Yale University.  
%-----------------------------------------------------------------------------------------------------------------------------------

%-----------------------------------------------------------------------------------------------------------------------------------
\bibliographystyle{mnras}
%\bibliography{bibliography2}

\bsp
%-----------------------------------------------------------------------------------------------------------------------------------

\label{lastpage}

\end{document}